\documentclass[12pt]{iopart}

\def\Diga{
\put(20,0){\circle*{2}} 
\qbezier(0, 0)(10, 15)(20, 0)
\qbezier(0, 0)(10, -15)(20, 0)\qquad\ 
}

\def\Digb{
\put(0,0){\circle*{2}}
\put(20,0){\circle*{2}}
\qbezier(0, 0)(10, 15)(20, 0)
\qbezier(0, 0)(10, -15)(20, 0)\qquad\  
}

\def\Figa{
\put(20,0){\circle*{2}}
\qbezier[30](0, 0)(10, 15)(20, 0)
\qbezier[30](0, 0)(10, -15)(20, 0)\qquad\  
}

\def\Figb{
\put(0,0){\circle*{2}}
\put(20,0){\circle*{2}}
\qbezier[30](0, 0)(10, 15)(20, 0)
\qbezier[30](0, 0)(10, -15)(20, 0)\qquad\  
}

\usepackage{graphicx}
\usepackage{pstricks}
\begin{document}

\title[An ab initio Calculation of the Universal Equation of State for the $O(N)$ Mode]
{An ab initio Calculation of the Universal Equation of State for the $O(N)$ Model}
\author{Denjoe O'Connor$^1$, J A Santiago$^2$ and C R Stephens$^3$}
\address{$^1$ School of Theoretical Physics,
Dublin Institute for Advanced Studies, 10 Burlington Road, Dublin 4, Ireland}
\address{$^2$Centro de Investigaci\'on en Matem\'aticas. 
Universidad Aut\'onoma del Estado de Hidalgo, Pachuca 42184,
M\'exico}
\address{$^3$ Instituto de Ciencias Nucleares, Universidad Nacional 
Aut\'onoma de M\'exico, Apartado Postal 70-543, M\'exico DF 04510, 
M\'exico}

\eads{\mailto{\mailto{denjoe@stp.dias.ie}, \mailto{sgarcia@uaeh.edu.mx}, 
\mailto{stephens@nucleares.unam.mx}}}

\begin{abstract}
Using an Environmentally Friendly Renormalization Group we derive an \textit{ab initio} universal
scaling form for the equation of state for the $O(N)$ model, $y=f(x)$, that
exhibits all required analyticity properties in the limits $x\rightarrow 0$, $x\rightarrow\infty$
and $x\rightarrow -1$. Unlike current methodologies based on a phenomenological scaling ansatz
the scaling function is derived solely from the underlying Landau-Ginzburg-Wilson 
Hamiltonian and depends only on the three Wilson functions $\gamma_\lambda$, $\gamma_\varphi$ 
and $\gamma_{\varphi^2}$ which exhibit a non-trivial crossover between the Wilson-Fisher 
fixed point and the strong coupling fixed point associated with the Goldstone modes on the 
coexistence curve. We give explicit results for $N=2$, 3 and 4 to one-loop order and 
compare with known results.
\end{abstract}



\section{Introduction}

The equation of state for the $O(N)$ model has been an object of intense scrutiny
over the last 30-40 years (see, for instance, \cite{zinnjustin} and \cite{Pelisseto} 
for recent reviews). It exhibits crossover behaviour between three distinct asymptotic 
regimes - the critical region approached along the critical isotherm, the critical region 
approached along the critical isochore, and, finally, the coexistence curve. It is the 
difficulty of encapsulating these distinct scaling behaviours within one overall scaling 
function that has prevented its \textit{ab initio} derivation from an underlying 
microscopic model. 

The first attempts at such an ab initio calculation, using the renormalization group 
(RG) and an $\varepsilon$-expansion \cite{Brezin}, foundered on the fact that they did
not exhibit Griffiths analyticity in the large $x=t/\varphi^{1/\beta}$ limit. This was 
due to the fact that the expansion was around a particular fixed point - the Wilson-Fisher 
fixed point. However, to obtain a universal equation of state, valid 
in the entire phase diagram, using the RG and \textit{without} phenomenological input, the
$\varphi$-dependent crossover between this fixed point and the strong coupling
fixed point associated with the massless Goldstone excitations on the coexistence curve 
must be accessed and controlled. In the latter regime the non-linear 
$\sigma$ model gives a good description \cite{Lawrie}. However, this model only accounts for the
Goldstone bosons and does not offer a full description of the phase diagram. 

As pure first principles calculations using a Landau-Ginzburg-Wilson Hamiltonian and the 
RG have not been capable of obtaining an equation of state valid
in all asymptotic regimes, the current ``state of the art'' \cite{zinnjustin,Pelisseto} is 
to base calculations on a parametrised phenomenological scaling form \cite{Schofield} that 
has the correct analyticity properties. Instead of the magnetisation, $\varphi$, reduced 
temperature, $t$, and magnetic field, $H$, new variables, $\theta$ and $R$, are introduced,
the relation between them being
\begin{eqnarray}
\varphi=m_0R^{\beta}m(\theta)\qquad\qquad t=R(1-\theta^2)\qquad\qquad H=h_0R^{\beta\delta}h(\theta)\nonumber
\end{eqnarray}
The particular functional dependence on $R$ ensures that Griffith's analyticity is preserved.
However, the two functions $m(\theta)$ and $h(\theta)$ are arbitrary. Hence, any particular 
choice constitutes a pure ansatz. Thus, Schofield's scaling form gives a large class of models, all of 
which, by construction, are consistent with Griffith's analyticity. 
How a particular microscopic model is related to this large class depends
on how these undetermined functions are represented, the canonical approach
being to represent them as polynomials. Thus, there is no unique map
between a microscopic model and a member of the class of Schofield parametrisations.

In this new parametrisation the scaling function of the equation of state
is given by
\begin{equation}
f\left( x\right)  =\left( \frac{m\left( \theta \right) }{m\left( 1\right) }%
\right) ^{-\delta }\frac{h\left( \theta \right) }{h\left( 1\right) }
\qquad{\rm where} \qquad
 x =\frac{1-\theta ^{2}}{\theta _{0}^{2}-1}\left( \frac{m\left( \theta _{0}\right) }
{m\left( \theta \right) }\right) ^{1/\beta }\ . 
\label{scot1} 
\end{equation}
Most current field theoretic formulations for determining the equation of
state (see \cite{Pelisseto} for a comprehensive review) rely on such
formulations. The drawback is that the underlying microscopic theory is not
used to determine the functional form of $m(\theta)$ and $h(\theta)$. Rather,
an ansatz is made as to the general functional form, which depends on certain
unknown parameters, and then the underlying microscopic theory is used to fix
them. The most common ansatz is that the functions are
polynomials in $\theta$. The coefficients of the powers of $\theta$ are
then determined by calculating the values of certain observables independently from the
underlying microscopic theory. 

In contrast to the above, in this paper\footnote{Which is a follow up to \cite{AnalyticIsing} 
which treated the case $N=1$}, using only the Landau-Ginzburg-Wilson Hamiltonian for the 
$O(N)$ model and by implementing an Environmentally Friendly RG \cite{Stephens2, Stephens1} 
that tracks the crossover between the fixed points that control the different asymptotic 
regimes, we derive an universal equation of state that obeys all required analyticity 
properties and where \emph{no} phenomenological input is required, only the three 
Wilson functions $\gamma_\varphi$, $\gamma_{\varphi^2}$ and $\gamma_\lambda$.

\section{The Equation of State}

In the critical region, the equation of state \cite{Widom} relates the external magnetic
field $H$, the reduced temperature $t$, and the magnetization $\varphi$,
\begin{equation}
y=f\left( x\right)  \label{fscaling}
\end{equation}
where the universal scaling function $f(x)$, is normalised such that $f(0)=1$ on the critical
isotherm, and $f(-1)=0$ on the coexistence curve. The scaling variables $y$ and $x$ are given by
$y=B_c^\delta H/\varphi^{\delta }$ and $x=B^{1/\beta}t/\varphi^{1/\beta }$.

The function $f(x)$ has expansions around the limits $x=0$ and $x\rightarrow\infty$ given by
\begin{equation}
f(x)=1+\sum_{n=1}^\infty f^0_n x^n\qquad\qquad 
\label{expansioninfty}
f(x)=x^\gamma\sum_{n=0}^\infty f^\infty_n x^{-2n\beta}
\end{equation}
In the limit $x\rightarrow\infty$ a natural variable is $z=b_1\varphi/t^{\beta}$, where 
$b_1$ an amplitude ratio in terms of which the equation of state takes the form
$H\propto t^{\beta\delta}F(z)$, where the universal scaling function $F(z)$ for 
small $z$ has an expansion of the form
\begin{equation}
F(z)=z +\frac{1}{6}z^3+\sum_{n=3}^\infty \frac{r_{2n}}{(2n-1)!}z^{2n-1}
\label{expansioninz}
\end{equation}
where $r_2=r_4=1$ by choice of normalisation.
As (\ref{expansioninz}) is an expansion in $\varphi$, the constants $r_{2n}$
are related to the $2n$-point correlation functions at $\varphi=0$ and hence
are very natural observables to calculate in lattice simulations. In the
limit $z\rightarrow\infty$, $F(z)$ has an expansion of the form
\begin{equation}
F(z)=z^\delta\sum_{k=0}^\infty F_k^\infty z^{-k/\beta}
\end{equation}
By relating $f(x)$ and $F(z)$
expansion coefficients of the two functions can be related to find
\begin{eqnarray}
f_n^\infty&=&z_0^{2n+1-\delta}\frac{r_{2n+2}}{F_0^\infty(2n+1)!}\label{fversusr}\\
f_n^0&=&\frac{F_n^\infty}{F_0^\infty}z_0^{-n/\beta}
\end{eqnarray}
where $z_0$ is the universal zero of the equation of state in terms of the variable $z$.
Thus, we see it is sufficient to know the expansion coefficients of $f(x)$ in
the limits $x\rightarrow 0$ and $\infty$ in order to calculate the asymptotic
properties of $F(z)$ and the interesting functions $r_{2n}$.  

Unlike the limits $x\rightarrow 0$ and $\infty$, near the coexistence curve,
$x\rightarrow -1$, there are no rigorous mathematical arguments as to the 
analyticity properties of $f(x)$, although there do exist conjectures. 
In \cite{Wallace}, based on an $\varepsilon$-expansion analysis, it was 
conjectured that $(1+x)$ has a double expansion in powers of $y$ and 
$y^{(d-2)/2}$ of the form  
\begin{equation}
1+x=c_{1}y+c_{2}y^{1-\epsilon /2}+d_{1}y^{2}+d_{2}y^{2-\epsilon /2}+...
\label{Coexi}
\end{equation}
In three dimensions it predicts an expansion of $(1+x)$ in powers of $y^{1/2}$.
Studies of the non-linear $\sigma$ model lead one to expect a leading behavior of the form
\begin{equation}
f(x)\sim c_f(1+x)^{2/(d-2)}
\end{equation}
though, as mentioned, the nature of the corrections to this behavior is not
well understood, although (\ref{Coexi}) is one conjecture. In the $1/N$
expansion there is some evidence \cite{Pelisseto} for logarithmic corrections
of the form $\ln(1+x)$ in three dimensions.

\section{A Renormalization Group Representation of the Equation of State}
\label{rgrepeqsta}

In this section we briefly outline the derivation of the equation of state
for a theory described by the standard Landau-Ginzburg-Wilson Hamiltonian with $O(N)$ symmetry
\begin{eqnarray}
{\cal H}[{\bf\varphi}]=\int\!
d^{d}x\left({1\over 2}\nabla\varphi^a\nabla\varphi^a
\!+{1\over 2}r(x)\varphi^a\varphi^a
\!+{\lambda_B\over 4!}(\varphi^a\varphi^a)^2\!\right)
\label{hamilt}
\end{eqnarray}
with $r=r_c+t_B$, where $r_c$ is the value of $r$
at the critical temperature $T_c$ and $t_B=\Lambda^2\ {(T-T_c)\over T_c}$,
$\Lambda$ being the microscopic scale.
Due to the Ward identities of the model, it is sufficient to know only the transversal 
correlation functions $\Gamma^{(N,M)}_t$, as all the other vertex functions can be
 reconstructed from these.  For instance, the equation of state itself is given by
$H=\Gamma^{(2)}_t\varphi$.


Due to the existence of large fluctations in the critical regime a
renormalization of the microscopic bare parameters of the form
\begin{eqnarray}
t(m_t,\kappa)&=&Z_{\varphi^2}^{-1}(\kappa)t_B(m_t)\label{coord1}\\
\lambda(\kappa)&=&Z_\lambda(\kappa)\lambda_B \label{coord2} \\
\varphi(\kappa)&=&Z_{\varphi}^{-1/2}(\kappa)\varphi_B \label{coord3}
\end{eqnarray}
must be imposed, where $\kappa$ is an arbitrary renormalization scale
and $m_t$ is the inverse transverse correlation length.
The renormalized parameters satisfy the differential equations
\begin{eqnarray}
\kappa{dt(\kappa)\over d\kappa}&=&\gamma_{\varphi^2}(\kappa)t(\kappa)\qquad{\rm where}\qquad
\gamma_{\varphi^2}(\kappa)=-{\kappa{d\over d\kappa}\ln Z_{\varphi^2}}\vert_c
\label{wilson1} \\
\kappa{d\lambda(\kappa)\over d\kappa}&=&\gamma_\lambda(\kappa)\lambda(\kappa)\qquad{\rm where}\qquad
\gamma_\lambda(\kappa)={\kappa{d\over d\kappa}\ln Z_{\lambda}}\vert_c
\label{wilson2} \\
\kappa{d\varphi(\kappa)\over d\kappa}&=&-{1\over2}\gamma_\varphi(\kappa)\varphi(\kappa)\qquad{\rm where}\qquad
\gamma_\varphi(\kappa)={\kappa{d\over d\kappa}\ln Z_{\varphi}}\vert_c
\label{wilson3}
\end{eqnarray}
where the right hand side are the Wilson functions associated with this coordinate transformation
and the derivative is taken along an appropriately chosen curve
in the phase diagram, which we here denote by $c$.
 
Integration of the RG equation for any multiplicatively renormalizable
$\Gamma^{(N,M)}_t$ yields
\begin{eqnarray}
\label{gammanmrge}
\Gamma^{(N,M)}_t(t,\lambda,\varphi)
=
{\rm e}^{\int^{m_t}_\kappa({N\over2}\gamma_\varphi-M\gamma_{\varphi^2}){dx\over
x}}
\Gamma^{(N,M)}_t(t(\kappa),\lambda(\kappa),\varphi(\kappa))
\end{eqnarray}
The renormalization constants $Z_\varphi$, $Z_{\varphi^2}$ and $Z_\lambda$ are fixed by 
imposing the explicitly magnetization dependent normalization conditions
on the transverse correlation functions
\begin{eqnarray}
\left.\partial_{p^2}\Gamma_t^{(2)}(p,t(\kappa,\kappa),\lambda(\kappa),
\varphi(\kappa),\kappa)\right\vert_{p^2=0}&=&1
\label{nctwof} \\
\Gamma_t^{(2,1)}(0,t(\kappa,\kappa),\lambda(\kappa),\varphi(\kappa),\kappa)
&=&1\label{ncthreef} \\
\Gamma_t^{(4)}(0,t(\kappa,\kappa),\lambda(\kappa),\varphi(\kappa),\kappa)
&=&\lambda .\label{ncfourf}
\end{eqnarray}
while the condition
\begin{eqnarray}
\kappa^2=\Gamma_t^{(2)}(0,t(\kappa,\kappa),\lambda(\kappa),\varphi(\kappa),\kappa),
\label{nconef}
\end{eqnarray}
serves as a gauge fixing condition that relates the sliding renormalization
scale $\kappa$ to the physical temperature $t$ and magnetization $\varphi$. 
Physically, $\kappa$ is a fiducial value of the non-linear scaling field $m_t$. 

Besides $m_t$, the other non-linear scaling field we use to
parametrise our results is
\begin{equation}
m_{\varphi}^2={1\over3}{\Gamma_t^{(4)}\varphi^2\over
\partial_{p^2}\Gamma_t^{(2)}\vert_{p^2=0}}
\label{mfdef}
\end{equation}
which is a RG invariant. It represents the anisotropy in the masses
of the longitudinal and transverse modes and is related to the
stiffness constant $\rho_s=\varphi^2\partial_{p^2}
\Gamma_t^{(2)}\vert_{p^2=0}$ via $m_{\varphi}^2= {1\over 3}\lambda\rho_s$.
With this renormalization prescription one may determine the equation
of state in terms of the non-linear scaling fields $m_t$ and $m_\varphi$,
as the transverse and longitudinal propagators that appear in all perturbative
diagrams can be parametrised in terms of them. 

\begin{figure}
   \centering
        \label{fig:univar:a}
   	\begin{pspicture}(0,0)(6.5,5.5)
           \psline{<->}(2.0,0.0)(2.0,5.0)	
           \psline{<->}(0.0,2.5)(6,2.5)		
           \rput(6.0,2.0){$T$}
           \rput(2.4,5.0){$\varphi$}
	   
%

  	\pscurve[linewidth=0.6mm](2.0,3.7)(4.0,2.5)(2.0,1.3)%
	\psline[linewidth=0.6mm](4.0,2.5)(5.5,2.5)
	\pscurve(2.0,4.3)(3.0,3.99)(4.0,3.45)(4.5,3.11)(4.9,3.0)(5.5,3.0)
	\pscurve(2.0,4.0)(3.0,3.69)(4.0,3.15)(4.5,2.81)(4.9,2.7)(5.5,2.7)

%
	\psdots[dotstyle=*,dotscale=1.3](3.0,3.39)(4.0,2.5)

%
  \psline[linestyle=dashed,linewidth=0.2mm]{->}(3.0,3.39)(4.5,3.39)
%
	\rput(4.3,2.0){$T_c$}
	\rput(3.0,3.0){$T_c(\varphi)$}
	
	\end{pspicture}
\caption{\small Integrating along curves of constant
$\varphi$ (or $t$) in the phase diagram starting from the coexistence curve.
}
\label{phasespace}
\end{figure}
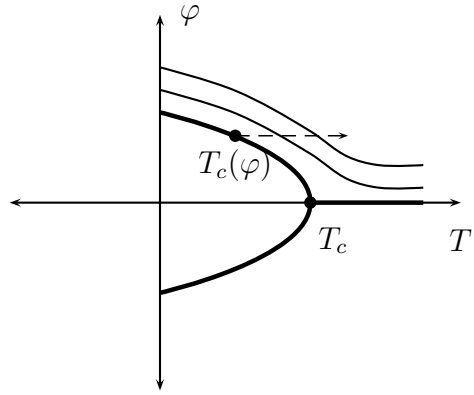

We now give a short review of the general methodology used to
derive the equation of state as developed in reference \cite{Stephens1,AnalyticIsing}. 
The equation can be found by integrating $d\Gamma_t^{(2)}(t,\varphi )$
along a curve of constant $\varphi$ (see figure (\ref{phasespace})), where in $dt= d\Gamma^{(2)}_t/
\Gamma^{(2,1)}_t$, the right hand side is written in the
coordinate system $(m_t,m_\varphi)$. 
We note the following points \cite{AnalyticIsing}:

1) We take $m_t=\kappa$ as the flow variable of the RG and hold $\varphi$ constant.
The non-linear scaling variable, $m_\varphi$, is then $\kappa$ dependent through 
its definition (\ref{mfdef}). 




2) We integrate $dt$ along a curve of constant $\varphi$ 
and fix the boundary condition on the coexistence curve, where $m_t=0$, whence we may write 
$(T-T_c(\varphi))= t + \Delta$, where $\Delta=(T_c-T_c(\varphi))$ 
is the temperature shift that measures the distance between the critical point 
and the point on the coexistence curve, $T_c(\varphi)$, see figure (\ref{phasespace}).
One finds  
\begin{eqnarray}
A_1(1+x)&=&\mathcal F(z)
\label{fundrelone}
\end{eqnarray}
where $z=m_t/m_\varphi$ and the scaling variable $x=B^{1/\beta}t/\varphi^{1/\beta}$, $B$ being a
amplitude. The universal scaling function $\mathcal F(z)$ is given by
\begin{eqnarray}
\mathcal F(z)=
\int_{0}^{z}\frac{2(2-\gamma _{\varphi })}{2-\gamma _{\lambda }
+\gamma _{\varphi }}%
e^{{-\int_{\infty }^{x}2
\left({\frac{\Delta \gamma _{\varphi ^{2}}-\frac{\Delta \gamma _{\lambda }%
}{2\beta }+\frac{\Delta \gamma _{\varphi }}{2\beta }}
{2-\gamma _{\lambda }+\gamma _{\varphi }%
}}\right) \frac{dy}{y}%
}}
x^{1\over\beta }\frac{dx}{x}\label{te}
\end{eqnarray}
where we have defined $\Delta\gamma_i=(\gamma_i-\gamma_i^{\rm WF})$, $WF$ denoting the 
Wilson-Fisher fixed point.
  
3) The equation of state is obtained using $H=\Gamma^{(2)}_t\varphi$ and the RG equation 
for $\Gamma_t^{(2)}$. One obtains $ H/\varphi^\delta =A_3^{-1}\, {\cal G}(z)$, with
the universal function
\begin{eqnarray}
{\cal G}(z) =
{\rm e}^{\frac{\gamma }{%
\beta }\int_{\infty }^{z}\frac{\Delta \gamma _{\lambda }-\Delta \gamma
_{\varphi }}{2-\gamma _{\lambda }+\gamma _{\varphi }}\frac{dy}{y}%
}{\rm e}^{-\int_{\infty }^{z}\frac{2\Delta \gamma _{\varphi }}
{2-\gamma _{\lambda}+\gamma _{\varphi }}\frac{dy}{y}}  
\end{eqnarray}
Defining the scaling variable $y=A_3^{-1}(H/\varphi^\delta)$ we then obtain the 
universal equation of state in the form (\ref{fscaling}) with
\begin{equation}
f(x)=\frac{1}{A_3}\mathcal G({\mathcal F}^{-1}(A_1(1+x)).
\end{equation}

4) The amplitudes $A_1$ and $A_3$ are both universal and can be determined as 
$A_1=\mathcal F(z_c)$ and  $A_3=\mathcal G(z_c)$, where $z_c$ is the value of $z$ 
that corresponds to the critical isotherm.

In order to determine the expansion coefficients $f_n^0$ and $f_n^\infty$,
as introduced in section 1, one requires the Taylor expansion of $f(x)$ around
$x=0$ and $x=\infty$. In terms of our parametric representation, $d^nf(x)/dx^n$ 
can be expressed using $d/dx=(dz/dx)d/dz$, where $dz/dx=A_1/({d\mathcal F(z)/dz})$ hence,
\begin{equation}
{d^nf(x)\over dx^n}={A_1\over A_3}
\left(\left({d\mathcal F(z)\over dz}\right)^{-1}{d\over dz}\right)^n\mathcal
G(z)
\end{equation}
which needs to be evaluated at the points of interest $z=z_c$ ($x=0$),
$z=\infty$ ($x=\infty$) and $z=0$ ($x=-1$).

\section{Results}

\subsection{The one loop Wilson functions}

To one loop order the bare vertex functions are given by
\begin{eqnarray}
\Gamma^{(2)}_t &=& p^2+r+{\lambda\over 6}\varphi^2 +{\lambda\over 2}\Diga +{\lambda\over 6}(N-1)\Figa\\
\Gamma^{(2, 1)}_t &=& 1-{\lambda\over 2}\left( \Digb +{N-1\over 3} \Figb\right)\\
\Gamma^{(4)}_t &=& \lambda -{3\over 2}\lambda^2\left(\Digb +{N-1\over 9}\Figb \right)
\end{eqnarray}
where the dashed line in the loops denotes the transversal propagator and the continuous line
the longitudinal one, i.e., $\Diga =\int {d^dq\over (2\pi)^d}  
{1\over q^2+t+{\lambda \over 2}\varphi^2}$ and
$\Figa=\int {d^dq\over (2\pi)^d} {1\over q^2+t+{\lambda\over 6}\varphi^2}$.     
Using (\ref{nctwof}), (\ref{ncthreef}) and (\ref{ncfourf}), the renormalization constants 
can be determined and hence the Wilson functions
\begin{eqnarray}
\gamma _{\lambda } &=&-\frac{3}{2}\lambda\kappa{d\over d\kappa}\left( \Digb +\frac{N-1}{9}
\Figb\right)\\
\gamma _{\varphi } &=& 0\\
\gamma _{\varphi ^{2}} &=&-\frac{\lambda }{2}\kappa{d\over d\kappa}\left( \Digb +\frac{N-1}{%
3}\Figb \right)  
\end{eqnarray}  

The running dimensionless coupling $\lambda$ satisfies
\begin{equation}
z\frac{d\lambda(z)}{dz}=-\varepsilon\lambda +c_d\lambda^2(z)
\left((1+{1\over z^2})^{d-6\over2}+{(N-1)\over9}\right)
\label{betafunction}
\end{equation}
Taking the initial condition
$\lambda(z_0)=\lambda$, in the limit $z_0\rightarrow\infty$,
$\lambda\rightarrow\infty$ one arrives at the universal separatrix
solution\footnote{This solution may also be reached by choosing the initial
coupling to be on the separatrix solution at $z=z_0$.}  
\begin{equation}
\lambda(z)=\left(c_d\left((1+{1\over z^2})^{d-6\over2}
+{(N-1)\over9}\right)\right)^{-1}
\end{equation}
On the separatrix
\begin{eqnarray}
\label{wilsononeloopon}
\gamma_\lambda&=&(4-d)\left({(1+{1\over z^2})^{d-6\over2}+{(N-1)\over9}\over
(1+{1\over z^2})^{d-4\over2}+{(N-1)\over9}}\right)\\
 \gamma _{\varphi ^{2}}&=&
(4-d)\left({(1+{1\over z^2})^{d-6\over2}+{(N-1)\over3}\over
3(1+{1\over z^2})^{d-4\over2}+{(N-1)\over3}}\right)\\
\gamma_\varphi&=&0
\end{eqnarray}

With the Wilson functions in hand we can calculate the scaling functions. In three dimensions
the scaling function $\mathcal G(z)$ is analytic, while the function $\mathcal F(z)$ can be written 
as an integral. Explicitly,
\begin{eqnarray}
\label{Gonelooparbitraryd}
\mathcal G(z)& =& z^4\prod_{i=1}^{3}\left( \sqrt{ 1+1/z^2 }- r_i\over 1-r_i \right)^{12\over 12+(N-1)r_i}\\
\mathcal F(z)&=&\int_0^z {4\over 2-\gamma_\lambda}\left( 1+\sqrt{ 1+1/x^2 }\over 2 \right)^{12(N-1)
\over (N-10)(N+8)}\nonumber\\
&\times &\prod_{i=1}^{3}\left( \sqrt{ 1+1/x^2 }- r_i\over 1-r_i \right)^{n_i}x^{20+N\over 8+N} \, dx 
\label{Fonelooparbitraryd}
\end{eqnarray}
where $r_i$ satisfies the cubic equation $(N-1)r_i^3+18r_i^2-9=0$ and 
$n_i={9(1-N)+(161+2N-N^2)r_i+9(N-1)r_i^2\over 12r_i+(N-1)r_i^2}{4\over (10-N)(N+8)}$. 

In the limit $z\rightarrow\infty$, which corresponds to approaching the
critical point along the critical isochore, the Wilson-Fisher fixed point is
approached and $\gamma_i\rightarrow\gamma^{\rm WF}_i$ with, at one loop,
$\gamma_\lambda=(4-d)$ and $\gamma _{\varphi ^{2}}=(4-d)(N+2)/(N+8)$.
On the contrary, in the limit $z\rightarrow 0$, which corresponds to approaching the
critical point along the coexistence curve $x\rightarrow -1$, the strong-coupling
fixed point is approached and $\gamma_i\rightarrow\gamma^{\rm SC}_i$.
There, the Goldstone bosons dominate and $\gamma_\lambda=\gamma _{\varphi ^{2}}=(4-d)$. 
Finally, the critical isotherm, $x=0$, is reached in the limit $z\rightarrow z_c$. 

\begin{figure}[h]
\begin{center}
\includegraphics[width=3.5in,angle=0]{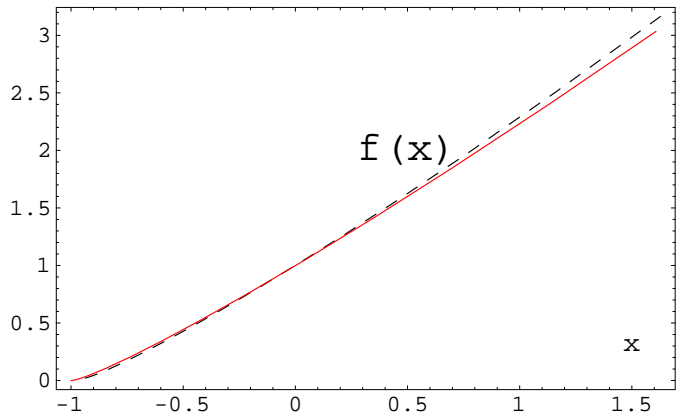}
\end{center}
\caption{\small The scaling function $f(x)$ for the
three-dimensional XY model. The dashed line was taken from reference \cite{Campostrini1}.
}   
\label{fscalinggraph1}
\end{figure}

\begin{figure}[h]
\begin{center}
\includegraphics[width=3.5in,angle=0]{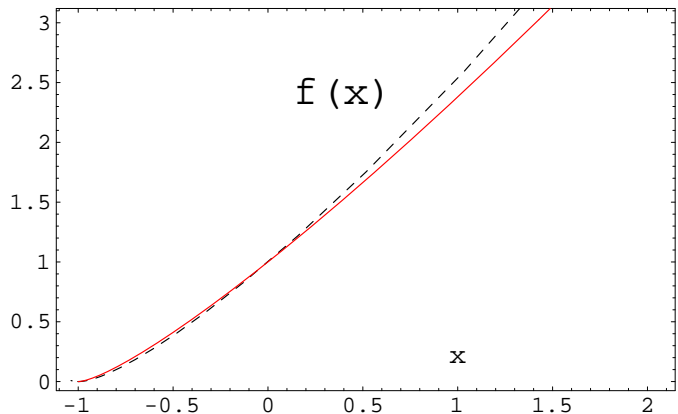}
\end{center}
\caption{\small The  scaling function $f(x)$ for the
three-dimensional Heisenberg model. The dashed line was taken from reference 
\cite{Campostrini2}.
}   
\label{fscalinggraph2}
\end{figure}

\begin{figure}[h]
\begin{center}
\includegraphics[width=3.5in,angle=0]{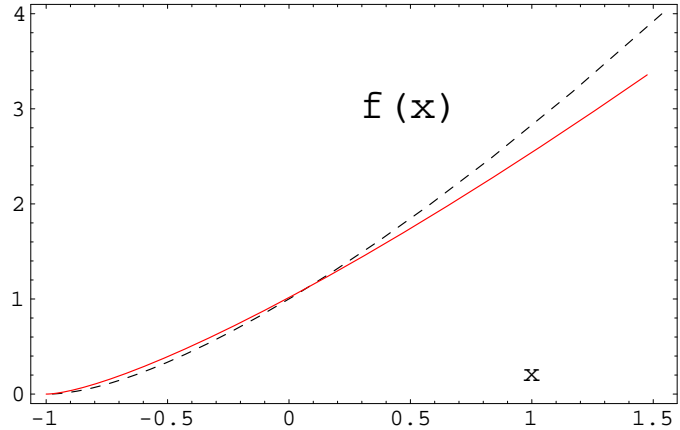}
\end{center}
\caption{\small The  scaling function $f(x)$ for the
three-dimensional $O(4)$ model. The dashed line was taken from reference \cite{Toldin}.
}   
\label{fscalinggraph3}
\end{figure}

With (\ref{Fonelooparbitraryd}) and (\ref{Gonelooparbitraryd}) we can plot the scaling function, $f(x)$,
for the full universal equation of state, as seen in Figures \ref{fscalinggraph1}, \ref{fscalinggraph2} and
\ref{fscalinggraph3} for the cases $N=2$, 3 and 4 respectively.

\subsection{The limit $z\rightarrow\infty$}

In the limit $z\rightarrow\infty$, the Wilson
functions can be expanded as power series in $z^{-2}$
\begin{equation}
\gamma_i(z)=\gamma^{WF}_i+\sum_{n=1}^\infty a_i(n)z^{-2n}
\end{equation}
Hence, the universal scaling functions $\mathcal F$ and $\mathcal G$
can also be written as power series in $z^{-2}$. This is true in a
diagrammatic expansion to all orders, not just at one loop. In the limit
$z\rightarrow 0$, $\gamma_i\rightarrow\gamma_i^{SC}$ but the nature of the corrections
is not obvious. At the one-loop level, from (\ref{wilsononeloopon}), one
can see that the leading corrections to the strong coupling
fixed point values will be $z^{(4-d)/2}$.  

To determine the constant $A_1$ we take the $z\rightarrow\infty$ limit of
(\ref{Fonelooparbitraryd}), identify the divergent part with $A_1x$ and the
constant remainder with $A_1$.\footnote{If $2(4-d)/3(d-2)$ is a positive
integer, $n$, then this remainder is zero and $A_1$ cannot be determined by looking at
the asymptotic limit $z\rightarrow\infty$. This is a pure artefact of the
one-loop approximation, where $\eta=0$, and has no physical meaning.}
For instance, for $N=1, 2,3,4$ we can write the divergent component of the function $\cal F$:
\begin{eqnarray}
I_1^\infty&=&4-4\left(1+{1\over 3}\sum_{i=1}^2 {1\over r_i -1}\right){1\over z^2}\\
I_2^\infty&=&4+\left(-{15\over 4}+{1\over 10}\sum_{i=1}^3 
{9-161r_i-9r_i^2\over r_i(r_i -1)(r_i+12)}\right){1\over z^2}\\
I_3^\infty&=&4+2\left(-{138\over 77}+{4\over 77}\sum_{i=1}^3 
{18-158r_i-18r_i^2\over r_i(r_i -1)(2r_i+12)}\right){1\over z^2}\\
I_4^\infty&=&4+\left( -{7\over 2} +{4\over 77}\sum_{i=1}^3{27-153r_i-27r_i^2\over r_i(r_i-1)(3r_i+12)}\right){1\over z^2}
\end{eqnarray}
The universal amplitudes $A_3$ can then be obtained using the corresponding equation
in 4) in section \ref{rgrepeqsta}. With these two constants, along with the two scaling functions 
$\mathcal F$ and $\mathcal G$ the scaling function of the universal equation of state may be 
obtained, and, correspondingly, any expansion coefficient, $f^0_n$, $f^\infty_n$ or $r_n$.

In Table \ref{U1} we see some results for a variety of expansion coefficients associated with 
the asymptotic regimes $x\rightarrow 0$ and $x\rightarrow\infty$. We also compare with various
other known results. The agreement is as good as one would expect from a one-loop approximation.
However, of note here is not the precision of the estimates but that they have been obtained
from an ab initio calculation where no phenomenological input was necessary. Thus, any expansion
coefficient is obtained by an expansion in the appropriate asymptotic limit of the universal
functions $\mathcal F(z)$ and $\mathcal G(z)$ which in their turn depend only on the Wilson functions.  
\begin{table}[h]
\caption{\label{U1}\small Values of some one-loop amplitudes for the 3-dimensional $O(N)$ model. 
We have written some known results as reported  in \cite{Pelisseto} (see table (21)) for N=2; 
\cite{Campostrini2} for N=3 and \cite{Toldin} for N=4.}
\begin{indented}
\item[]\begin{tabular}{@{}llll}
\br
$N$ & $2$ & $3$ & $4$  \\
\mr
\hline\hline          
$z_c$   & $0.694$   &  $0.677$  & $0.656$    \\

\hline\hline

$A_1$   &  $0.959$   & $0.934$   & $0.897$    \\

\hline\hline

$A_3$   & 1.022   & 0.840      & 0.698   \\

\hline\hline

$R^4_+$   & 6.35 ( 7.6(2) )  & 6.06 (7.8(3) )    & 8.86  ( 7.6(4) )   \\

\hline\hline

$F_0^\infty$   & 0.035 ( 0.0304(3) )  & 0.0348 (0.0266(5))    & 0.015 ( 0.0240(5) )  \\

\hline\hline

$R_\chi$   & 1.42 (1.41(6))   & 1.280 (1.31(7))     & 1.18 ( 1.12(11) )  \\

\hline\hline

$f_1^\infty$ & 0.746 & 0.789 & 1.246 \\

\hline\hline

$f_2^\infty$ & 0.641 & 0.695 & 1.171 \\

\hline\hline

$f_3^\infty$ & 0.103 & 0.097 &  0.151 \\

\hline\hline

$f_4^\infty$ & -0.005 & -0.006 &  -0.014 \\

\hline\hline

$c_f$   & 15.6 (15(10))  & 4.92 ( 5(3))     & 2.66 ( 2.8(1.4) )  \\

\hline\hline

$f_1^0$   &  1.16   &  1.26 (1.34(5))      & 1.35 ( 1.5(8))   \\

\hline\hline

$f_2^0$   &  0.086   &  0.143 (0.020(2))     &  0.2031 ( 0.33(5))   \\

\hline\hline

$f_3^0$   &  -0.022   &  -0.040(-0.10(1))      & -0.056 (-0.08(2) )  \\

\hline\hline

$r_6$   &  2.704 ( 1.951(14) )  &  2.90 ( 2.1(6))  & 2.11 ( 1.79(2) )   \\

\hline\hline

$r_8$   &  2.873 ( 1.36(9) )   &  2.80 ( 0.6(2) )     &  1.29   ( 0.2(4)) \\

\hline\hline

$r_{10}$   &  -1.58 ( -7(5) )   & - 2.05 ( -6(3) )     &  -0.97   ( -5(6)) \\

\br
\end{tabular}
\end{indented}
\end{table}

\subsection{The limit $z\rightarrow 0$}

In the limit $z\rightarrow 0$, corresponding to the coexistence curve, the parametric functions 
${\cal F}(z)$ and ${\cal G}(z)$ can be written as power series in the scaling variable $z$ as
${\cal F}(z)=A_1(1+x)={a\over 2} z^2+{b\over 3}z^3+{c\over 4}z^4+\dots $ and 
${\cal G}(z)=A_3\, y= d z^4+f z^5+gz^6+\dots$. We can invert the equation for $\mathcal F(z)$ 
to obtain the non-linear scaling field $z$ in terms of $(1+x)$ and then, once we substitute into the function 
${\cal G}(z)$, we obtain the equation of state near the coexistence curve in the form
\begin{equation}
y=c_f(1+x)^2+c_1(1+x)^{5/2}+c_2(1+x)+\dots
\end{equation}
or alternatively 
\begin{equation}
(1+x)=\tilde c_1 y + \tilde c_2  y^{1/2} +\tilde c_3 y^{3/4}+\dots
\end{equation}
In Table \ref{U2} we give the first expansion coefficients and compare them to the results of \cite{Wallace}.
\begin{table}[h]
\caption{\label{U2}\small Values of some one-loop amplitudes for the 3-dimensional $O(N)$ model in the limit $z\rightarrow 0$. 
The values in parenthesis for $\tilde c_1$ y $\tilde c_2$ were taken from \cite{Wallace}, where an $\varepsilon$
expansion was used.}
\begin{indented}
\item[]\begin{tabular}{@{}llll}
\br
$N$ & $2$ & $3$ & $4$  \\
\mr

\hline\hline

$c_1$   &  $ -111.6 $   & $ -17.8  $   & $ -6.4  $    \\

\hline\hline

$c_2$   & $ 399.8 $  & $ 32.0 $      & $ 7.6$   \\

\hline\hline

${\tilde c_1}$   & $ 0.82 (0.9)  $  &  $ 0.66 (0.82  )  $    &  $  0.54 ( 0.75)  $   \\

\hline\hline

${\tilde c_2}$   & $ 0.25 ( 0.1)  $ &  $ 0.45 (0.18 ) $   & $ 0.61 (0.25 ) $  \\

\hline\hline

${\tilde c_3}$   & $ 0.46  $ & $0.55   $    & $0.57  $  \\

\br
\end{tabular}
\end{indented}
\end{table}

\section{Conclusions}

Although there exist established methods for calculating the universal equation of 
state, $f(x)$, while preserving
Griffiths analyticity, they are based on a \emph{phenomenological} scaling 
ansatz that has no underlying microscopic basis. On the other hand it has not
been possible to preserve all asymptotic properties of $f(x)$ starting from a 
Landau-Ginzburg-Wilson Hamiltonian using the RG, due to the 
fact that the latter involved an expansion around the Wilson-Fisher fixed point.
In order to calculate $f(x)$ using RG methods it is necessary 
to be able to capture $\varphi$ dependent crossover between the two different
fixed points. 

In this paper we have developed a RG that captures the 
crossover between the Wilson-Fisher and strong-coupling fixed points and therefore
captures all elements of Griffiths analyticity, accessing all three different 
scaling regimes. Thus, we have achieved an ab initio calculation of the universal
equation of state from an underlying microscopic model with no phenomenological
input. The calculation depends only on the three Wilson functions $\gamma_\varphi$,
$\gamma_{\varphi^2}$ and $\gamma_\lambda$. We used the method to calculate $f(x)$
to one loop for arbitrary $N$ and made a comparison to known results from other
methods. The novelty of the present approach is not in the precision of any
result for an expansion coefficient, as here we have only worked to one loop, 
but rather in exhibiting a method that order by order in perturbation theory
preserves all necessary analyticity properties of $f(x)$. However, the method   
can be extended to higher loop order. The chief difficulty in doing so is that
the higher order Feynman diagrams must be evaluated numerically (remembering that
they are crossover functions not constants) as there are no closed form expressions
for them. As the amplitude $A_1$ involves cancelling two divergent expressions
this also is trickier when done numerically. We will return to the question of 
two-loop calculations in a future publication.

\ack
CRS and JAS would like to thank DIAS Dublin for hospitality while some of this work was 
carried out. JAS aknowledges CONACyT Grant 53187-F and PROMEP for financial support.

\end{document}